\documentclass[12pt]{article}

\usepackage{amsmath}
\usepackage{amsfonts}
\usepackage{amssymb}
\usepackage{graphicx}
\usepackage{subfigure}
\usepackage{rotating}
\usepackage{color}
\usepackage{natbib}
\usepackage{graphicx,setspace,amsmath,amssymb,subfigure,url,multirow,booktabs,
verbatim,bm}

\allowdisplaybreaks[3]

\newtheorem{theorem}{Theorem}

\newtheorem{remark}{Remark}
\newtheorem{corollary}{Corollary}

\def\bse{\begin{eqnarray*}}
\def\ese{\end{eqnarray*}}
\def\beq{\begin{equation}}
\def\eeq{\end{equation}}
\def\bqe{\begin{eqnarray}}
\def\eqe{\end{eqnarray}}

\def\boldr{{\mbox{\boldmath $r$}}}
\def\boldu{{\mbox{\boldmath $u$}}}
\def\boldnu{{\mbox{\boldmath $\nu$}}}
\def\boldC{{\mbox{\boldmath $C$}}}
\def\boldT{{\mbox{\boldmath $T$}}}
\def\boldX{{\mbox{\boldmath $X$}}}
\def\boldy{{\mbox{\boldmath $y$}}}
\def\boldY{{\mbox{\boldmath $Y$}}}
\def\boldepsilon{{\mbox{\boldmath $\epsilon$}}}
\def\bolddelta{{\mbox{\boldmath $\delta$}}}
\newcommand{\bbeta}{\boldsymbol{\beta}}
\def\diag{{\mbox{\rm diag}}}

\usepackage{setspace}

\begin{document}

\begin{spacing}{1.0}


\thispagestyle{empty}

\title{\bf Penalized integrative analysis under the accelerated failure time model}

\author{
Qingzhao Zhang$^1$, Sanguo Zhang$^1$, Jin Liu$^2$, Jian Huang$^3$ and
Shuangge Ma$^{4*}$\\ \\
$^1$School of Mathematical Sciences, University of Chinese Academy of Sciences\\
$^2$Center of Quantitative Medicine, Duke-NUS School of Medicine\\
$^3$Department of Statistics and Actuarial Science, University of Iowa\\
$^{4}$Department of Biostatistics, Yale University\\
{\it *email: shuangge.ma@yale.edu}}
\maketitle

{\bf Running Title}: Integrative analysis under AFT model

\begin{abstract}
For survival data with high-dimensional covariates, results generated in the analysis of a single dataset are often unsatisfactory because of the small sample size. Integrative analysis pools raw data from multiple independent studies with comparable designs, effectively increases sample size, and has better performance than meta-analysis and single-dataset analysis. In this study, we conduct integrative analysis of survival data under the accelerated failure time (AFT) model. The sparsity structures of multiple datasets are described using the homogeneity and heterogeneity models. For variable selection under the homogeneity model, we adopt group penalization approaches. For variable selection under the heterogeneity model, we use composite penalization and sparse group penalization approaches. As a major advancement from the existing studies, the asymptotic selection and estimation properties are rigorously established. Simulation study is conducted to compare different penalization methods and against alternatives. We also analyze four lung cancer prognosis datasets with gene expression measurements.
\end{abstract}


\section{Introduction}
In survival studies, data with high-dimensional covariates are now commonly encountered. A lung cancer prognosis study with gene expression measurements is presented in this article, and more are available in the literature. With such ``large $p$, small $n$" data, results generated in the analysis of a single dataset are often unsatisfactory because of the small sample size \citep{guerra, liu2013sparse, ma2011integrative2}. For outcomes of common interest, there are often multiple independent studies with comparable designs. This makes it possible to pool multiple datasets, increase sample size, and improve over single-dataset analysis. As a family of multi-dataset analysis methods, integrative analysis methods pool and analyze raw data from multiple studies and outperform classic meta-analysis methods, which analyze multiple datasets separately and then combine summary statistics.

In this article, we conduct the integrative analysis of multiple independent survival datasets under the accelerated failure time (AFT) model. The analysis goal is to identify, out of a large number of measured covariates, important markers associated with survival. For such a purpose, we adopt penalization, which has been the choice of many high-dimensional studies. A large number of penalization methods have been developed for single-dataset analysis. However because of the multi-dataset settings and heterogeneity across datasets, they are not applicable to integrative analysis. The sparsity structures of multiple datasets can be described using the homogeneity and heterogeneity models. Different models demand marker selection with different properties and hence different methods. This makes integrative analysis even more complicated. Penalization methods for integrative analysis have been developed \citep{liu2013sparse, ma2011integrative2}, however, in an unsystematic manner.

This study advances from the existing ones in the following aspects. First, it advances from single-dataset analysis and meta-analysis by conducting integrative analysis of multiple heterogeneous datasets. Second, it conducts more systematic investigation than the existing integrative analysis studies such as \cite{liu2013sparse, ma2011integrative2}. {\it More importantly, it rigorously establishes the selection and estimation properties which have not been previously examined.} The theoretical development is nontrivial because of data complexity, model settings, and penalties. Third, the properties of composite penalization and sparse group penalization have not been studied for single-dataset analysis under the AFT model. Thus our study can also provide insights for single-dataset penalization methods. Fourth, this study also advances from the existing studies by conducting systematic simulations and direct comparisons of multiple methods.

Data and model settings are described in Section 2. Penalized integrative analyses under the homogeneity and heterogeneity models are investigated in Section 3 and 4 respectively. We conduct numerical study in Section 5. The article concludes with discussions in Section 6. Technical details and additional analysis results are provided in Appendix.

\section{Integrative analysis under AFT model}

Consider the integrative analysis of survival data from $M$ independent studies. In study $m(=1, \ldots, M)$ with $n_m$ iid subjects, let $\boldT^m =(T_1^m,\cdots,T_{n_m}^m)^\top$ be the logarithm of failure times and $\boldX^m\in R^{n_m\times p_m}$ be the predictor matrix.  Assume the AFT model
\bqe
\boldT^m = \boldX^m\bbeta^m+\boldepsilon^m.
\eqe
$\bbeta^m$ is the vector of regression coefficients, and $\boldepsilon^m$ is the vector of random errors. With proper normalization, the intercept term has been omitted. Assume that all datasets measure the same set of covariates. Then $p_1=\cdots=p_M=p$. When different datasets have mismatched covariate sets, a rescaling approach \citep{ma2011integrative1, liu2013sparse} can be adopted. The proposed approaches are then applicable with minor modifications.

Let $\bbeta=(\bbeta^1,\cdots,\bbeta^M)=(\bbeta_1,\cdots,\bbeta_p)^\top$, where $\bbeta_j=(\beta_j^1,\cdots,\beta_j^M)^\top$ consists of the coefficients of
variable $j$ in all $M$ datasets. Moreover, write $\bbeta = (\beta_{ij})_{p\times M}$ with its true value $\bbeta^*$, where $\beta_{ij}=\beta_i^j$. With the heterogeneity across datasets, $\beta_j^m$ is not necessarily equal to $\beta_j^{k}$ for $m \neq k$. Under right censoring, one observes $(\boldY^m, \bolddelta^m, \boldX^m)$ with $\boldY^m =\boldT^m \wedge \boldC^m$, where $\boldC^m$ is the vector of log censoring times, and $\bolddelta^m=1\{\boldT^m\leq \boldC^m\}$.

When the distribution of random errors is unknown, there are multiple estimation approaches \citep{ying1993large}. We adopt the weighted least squares (LS) approach \citep{stute1993consistent}, which has the lowest computational cost and is desirable with high-dimensional data.
Let $\hat{F}^m$ be the Kaplan-Meier estimator of the distribution function $F^m$ of $T^m$. Let $Y_{(1)}^m\leq \cdots \leq Y_{(n_m)}^m$ be the order statistics of $Y_i^m$'s. $\hat{F}^m$ can be written as $\hat{F}^m(y)=\sum_{i=1}^{n_m} \omega_{i}^m 1\{Y_i^m\leq y\}$, where $\omega_{i}^m$'s are expressed as
$\omega_{1}^m = \frac{\delta_{(1)}^m}{n_m}$ and $\omega_{i}^m = \frac{\delta_{(i)}^m}{n_m-i+1}\prod_{j=1}^{i-1}
\left(\frac{n_m-j}{n_m-j+1}\right)^{\delta_{(j)}^m}, i=2,\cdots, n_m.$
Here $\delta_{(1)}^m, \cdots, \delta_{(n_m)}^m$ are the associated censoring indicators of the ordered $Y_i^m$'s. Denote $W_m =\diag\{n_m\omega_{1}^m, \cdots, n_m\omega_{n_m}^m\}$. Then for the $M$ datasets combined, the weighted LS approach is to minimize
\bqe\label{obj0}
\tilde{L}(\bbeta)=\frac{1}{2n}\sum_{m=1}^M (\boldY^m- \boldX^m\bbeta^m)^\top W_m (\boldY^m- \boldX^m\bbeta^m).
\eqe
Note that the components of $\boldY^m$ and $\boldX^m$ need to be sorted. Assume that:

\noindent[{\sc Condition 1}]~ (a)~The $n_m$ components of $\boldepsilon^m$ are i.i.d. and sub-Gaussian with noise level $\sigma_m$. That is, for all vector $\boldnu$ with $\|\boldnu\|_2=1$ and any $t\geq 0$, $P(|\boldnu^\top\boldepsilon^m|\geq t)\leq 2 \exp\left(-\frac{t^2}{2\sigma_m^2}\right)$. (b) $\boldepsilon^m$ is independent of $W_m$.

The total sample size is $n=\sum_{m = 1}^M {n_m}$. The important predictor index sets of $M$ datasets are respectively labeled as $S_1, \cdots, S_M$. Then $S = \bigcup\limits_{m=1}^M S_m$ denotes the important set with its corresponding variables important in at least one dataset. Let $S^c$ and $|S|$ denote the complement and cardinality of set $S$, respectively. Let $\mathcal {A}=\{(i,j): \beta_{ij}^*\neq 0\}$ and $\mathcal {B}=\{(i,j): i\in S , j=1,\cdots,M\}$. Let $\bbeta_\mathcal {A}$ and $\bbeta_\mathcal {B}$ denote the components of $\bbeta$ indexed by $\mathcal {A}$ and $\mathcal {B}$, respectively. For a $p\times 1$ vector $v$ and index set $I\subset \{1,\cdots,p\}$, let $v_I$ denote the components of $v$ indexed by $I$. Moreover, let $\boldX^{m,i}$ denotes the transposition of the $i$th row of $\boldX^m$. Then for any index set $I\subset \{1,\cdots,p\}$,
$\boldX_I^m = \left(\boldX_I^{m,1},\cdots,\boldX_I^{m,n_m}\right)^\top.$

\subsection{Homogeneity and heterogeneity models}

The sparsity structure of $\bbeta$ can be described using the homogeneity and heterogeneity models.
Under the homogeneity model, $\bbeta^m$'s have the same sparsity structure. That is, $I(\beta^m_j=0)=I(\beta^k_j=0)$ for all $(m,k,j)$'s. The intuition is that if the $M$ datasets are ``close enough", then the same set of markers should be identified in all datasets. Under this model, we only need to determine whether a covariate is important or not, that is, only one level of selection is needed. With the (sometimes great) differences across datasets, the homogeneity model may be too restricted. As an alternative, the heterogeneity model allows different datasets to have different sparsity structures. It includes the homogeneity model as a special case and can be more flexible. Under this model, we need to determine whether a covariate is associated with any response at all. In addition, for an important covariate, we need to determine in which datasets it is important. That is, a two-level selection is needed.

\section{Integrative analysis under the homogeneity model}

Under this model, one-level selection is needed and can be achieved using group penalization. In terms of formulation and computation, the development of group penalization methods in integrative analysis share some similarity with that in single-dataset analysis \citep{buhlmann2011statistics}. However, with the significantly different data settings and adoption of the AFT model, the theoretical development has significant differences.

\subsection{Group LASSO}
Consider the group LASSO penalized objective function
\bqe\label{obj1}
L(\bbeta)=\frac{1}{2n}\sum_{m=1}^M (\boldY^m- \boldX^m\bbeta^m)^\top W_m (\boldY^m- \boldX^m\bbeta^m)+\lambda\sum_{j=1}^p \|\bbeta_j\|_2,
\eqe
where $\lambda$ is the tuning parameter and $\|\bbeta_j\|_2=\left[(\beta_j^1 )^2+\cdots+(\beta_j^M)^2\right]^{1/2}$.

For set $S$, define the estimate $\hat{\bbeta}_\mathcal {B}=(\hat{\bbeta}_S^1,\cdots,\hat{\bbeta}_S^M)$ as
\bqe\label{oralasso}
\hat{\bbeta}_\mathcal {B}={\arg\min}_{\bbeta_\mathcal {B}}\left\{\frac{1}{2n}\sum_{m=1}^M (\boldY^m- \boldX_S^m\bbeta_S^m)^\top W_m (\boldY^m- \boldX_S^m\bbeta_S^m)+\lambda \sum_{j\in S} \|\bbeta_j\|_2\right\}.
\eqe
For group LASSO to be able to consistently identify the true sparsity structure, there needs a local solution  $\hat{\bbeta}^{glasso}=\{\hat{\bbeta}^{glasso}_\mathcal {B}, \hat{\bbeta}^{glasso}_{\mathcal {B}^c}\}$ for (\ref{obj1}), where $\hat{\bbeta}^{glasso}_\mathcal {B}=\hat{\bbeta}_\mathcal {B}$ and $\hat{\bbeta}^{glasso}_{\mathcal {B}^c}=0$.
Define $$\bar{\rho}_{2}^m = \lambda_{\max}\{n_m^{-1}\boldX_{S}^{m\top}{ W_m} ^2 \boldX^m_{S}\}, ~\underline{\rho}_{1}^m  = \lambda_{\min}\{n_m^{-1}\boldX_{S}^{m\top} { W_m} \boldX^m_{S}\}$$
$$\Lambda_m= \max_{j}\{n_m^{-1}\boldX_j^{m\top}{ W_m} ^2 \boldX_j^m\},~ \psi_m = \|\boldX_{S^c}^{m\top}{ W_m}\boldX^m_{S}(\boldX_{S}^{m\top} { W_m}\boldX^m_{S})^{-1}\|_\infty.$$
\begin{theorem}\label{lasso}
Consider the estimator defined by minimizing (\ref{obj1}). Under Condition 1,
\begin{enumerate}
  \item  There exists a local minimizer $\hat{\bbeta}_\mathcal {B}$ of (\ref{oralasso}) such that
  \[\Pr\left\{\|\hat{\bbeta}_S^m - {\bbeta_{S}^m}^*\|_2\leq \lambda\sqrt{|S|} \frac{4}{\underline{\rho}_{1}^m }\frac{n}{n_m}, m=1,\cdots,M\right\}\geq 1-\sum\limits_{m=1}^M \exp\left(-\frac{\lambda^2 |S| n^2}{2 \sigma_m^2 \bar{\rho}_{2}^m n_m  }\right).\]
  \item Assume the ir-representable conditions $\psi_m \leq D_m<1$.  $\hat{\bbeta}^{glasso}=\{\hat{\bbeta}^{glasso}_\mathcal {B}, \hat{\bbeta}^{glasso}_{\mathcal {B}^c}\}$ with $\hat{\bbeta}^{glasso}_\mathcal {B}=\hat{\bbeta}_\mathcal {B}, \hat{\bbeta}^{glasso}_{\mathcal {B}^c}=0$ is a local minimizer of (\ref{obj1}) with probability at least \[1-\sum\limits_{m=1}^M \exp\left(-\frac{\lambda^2 |S|n^2}{2 \sigma_m^2 \bar{\rho}_{2}^m n_m  }\right)-2p \sum\limits_{m=1}^M\exp\left\{-\frac{n^2\lambda^2(1-D_m)^2}{2n_m \Lambda_m\sigma_m^2(1+D_m)^2}\right\}.\]
\end{enumerate}
\end{theorem}
In single-dataset analysis, \cite{zhao2006model} and followup studies establish selection consistency under the ir-representable condition. Under a similar condition for individual datasets, integrative analysis also has selection consistency.

With the probability bounds in Theorem 1, we can obtain a more straightforward understanding of the penalized estimators and derive the following result.
\begin{corollary}\label{corollary-1}
Suppose that for $m=1,\cdots, M$, $\underline{\rho}_{1}^m, \bar{\rho}_{2}^m$, and $\Lambda_m$ are bounded away from zero and infinity. Assume that $n/n_m = O(1)$, $|S|\ll n$, and $\log p = O(n^\alpha)$ with $\alpha<1$. Under Condition 1 and the ir-representable conditions in Theorem 1, if
$ |S|^{-1/2}\min\limits_{j\in S}\|\bbeta_j^*\|_2\gg\lambda \gg n^{\frac{\alpha-1}{2}}$, then group LASSO can identify the true sparsity structure and $\|\hat{\bbeta}_S^m - {\bbeta_{S}^m}^*\|_2 = O_p(\lambda\sqrt{|S|})$, $m=1,\cdots, M$.
\end{corollary}

\begin{remark}
It is known that in single-dataset analysis the group LASSO is group selection consistent under some variants of the ir-representable condition. See \cite{huang2012selective} and others for reference. Similar conditions are needed in the integrative analysis with group LASSO. The conditions in Corollary 1 on $\underline{\rho}_{1}^m, \bar{\rho}_{2}^m $, and $\Lambda_m$ are on the design matrixes and censoring probabilities. Corollary 1 shows that even when the group LASSO can identify the true sparsity structure, $\lambda$ should be much large than $n^{-1/2}$, leading to $\|\hat{\bbeta}_S^m - {\bbeta_{S}^m}^*\|_2 \gg \sqrt{|S|/{n}}$.
\end{remark}

\subsection{Concave 2-norm group selection}
Consider penalization built on concave penalties. Notable examples of concave penalty include SCAD \citep{fan2001variable} and MCP \citep{zhang2010nearly}. For $t\geq 0$, the SCAD penalty has first order derivative $p'_\lambda(t) = \lambda \left\{I(t\leq \lambda)+\frac{(a\lambda-t)_+}{(a-1)\lambda}I(t>\lambda)\right\}$, for some $a>2$. The MCP has derivative
$p'_\lambda(t) = \lambda\left(1-\frac{t}{a\lambda}\right)_+$, for some $a>1$. Consider the objective function
\bqe\label{obj2}
L(\bbeta)=\frac{1}{2n}\sum_{m=1}^M (\boldY^m- \boldX^m\bbeta^m)^\top W_m (\boldY^m- \boldX^m\bbeta^m)+\sum_{j=1}^p p_\lambda(\|\bbeta_j\|_2),
\eqe
where the penalty $p_\lambda(\cdot)$ satisfies:

\noindent
[{\sc Condition 2}] $\lambda ^{-1}p_\lambda(t)$ is concave in $t\in[0,\infty)$ with a continuous derivative $\lambda ^{-1} p'_\lambda(t)$ satisfying $\lambda ^{-1} p'_\lambda(0+)\in(0,\infty)$. In addition, $\lambda ^{-1}p'_\lambda(t)$ is increasing in $\lambda\in(0,+\infty)$, and $\lambda ^{-1} p'_\lambda(0+)$ is independent of $\lambda$.

\noindent
[{\sc Condition 3}] $\theta=\inf\left\{\frac{t}{\lambda}:~\lambda^{-1}p'_\lambda(t)=0, ~t\geq 0\right\}$ is bounded.
\begin{remark}
Condition 2 is also considered by \cite{fan2011nonconcave}. LASSO, SCAD, and MCP all satisfy this condition. Condition 3 is added to guarantee unbiasedness. LASSO does not satisfy Condition 3 since $\lambda^{-1}p'_\lambda(t)=1$ leads to $\theta=\infty$, while SCAD and MCP satisfy with $\theta = a$. Another approach that has been studied is the 2-norm group bridge \citep{ma2012identification}. Under certain conditions, its selection consistency is established in \cite{ma2011integrative1}. Note that the bridge penalty does not satisfy Condition 3 and needs to be separately investigated.
\end{remark}

Consider the properties of concave 2-norm group penalization. Define the oracle estimator as $\hat{\bbeta}^{oracle}=\{\hat{\bbeta}^{oracle}_\mathcal {B}, \hat{\bbeta}^{oracle}_{\mathcal {B}^c}\}$ with $\hat{\bbeta}^{oracle}_\mathcal {B}=\tilde{\bbeta}_\mathcal {B}$ and $\hat{\bbeta}^{oracle}_{\mathcal {B}^c}=0$, where
\bqe\label{oracle1}
\tilde{\bbeta}_\mathcal {B} = {\arg\min}_{\bbeta_S}\left\{\frac{1}{2n}\sum_{m=1}^M (\boldY^m- \boldX_S^m\bbeta_S^m)^\top W_m (\boldY^m- \boldX_S^m\bbeta_S^m)\right\}.
\eqe
\begin{theorem}\label{scad}
Under Condition 1-3, consider the estimator defined by minimizing (\ref{obj2}).
\begin{enumerate}
  \item For any $R_m<\sqrt{\frac{n_m}{|S|}}$, we have
  $$\Pr\left(\|\tilde{\bbeta}_S^m-{\bbeta_S^m}^*\|\leq \sqrt{\frac{|S|}{n_m}}R_m,~m=1,\cdots,M\right)
\geq 1-\sum\limits_{m = 1}^M \exp\left\{-R_m^2\frac{|S|(\underline{\rho}_{1}^m )^2}{8\bar{\rho}_2^m\sigma_m^2 }\right\}.$$
  \item  Suppose $\lambda < \frac{\min\limits_{j\in S}\|\bbeta_j^*\|_2}{2\theta}$ and $R_m^\dag \leq \frac{\min\limits_{j\in S}\|\bbeta_j^*\|_2}{2\sqrt{M}}\sqrt{\frac{n_m}{|S|}}$. Then with probability at least \[1- \sum\limits_{m = 1}^M \exp\left\{-\frac{|S|(\underline{\rho}_{1}^m )^2}{8\bar{\rho}_2^m \sigma_m^2 }R_m^{\dag 2}\right\}-2p \sum\limits_{m=1}^M\exp\left\{-\frac{n^2 p'^2_\lambda(0+)}{2n_m \Lambda_m \sigma_m^2(1+\psi_m)^2}\right\},\]
    $\hat{\bbeta}^{oracle}$ is a local minimizer of (\ref{obj2}).
\end{enumerate}
\end{theorem}
Theorem 2 can be used to derive the following asymptotic result.
\begin{corollary}\label{corollary-2}
Suppose that for $m=1,\cdots, M$, $\underline{\rho}_{1}^m, \bar{\rho}_{2}^m $ and $\Lambda_m$ are bounded away from zero and infinity, $n/n_m = O(1)$, $|S|\ll n$, $\log p = O(n^\alpha)$ with $\alpha<1$, and $\psi_m = O(n^{\alpha_1})$ with $\alpha_1\in[0,1/2)$. Under Condition 1-3, if
$\lambda < \frac{\min\limits_{j\in S}\|\bbeta_j^*\|}{2\theta}$ and $\lambda \gg n^{\frac{\alpha-1}{2}+\alpha_1}$, then the concave 2-norm group selection can identify the true sparsity structure and $\|\hat{\bbeta}_S^m - {\bbeta_{S}^m}^*\|_2 = O_p(\sqrt{\frac{|S|}{n_m}})$.
\end{corollary}
\begin{remark}
When the concave penalty is used, the upper bound of $\psi_m$ can grow to $\infty$ at rate $O(n^{\alpha_1})$. In contrast, the group LASSO needs the ir-representable conditions. Moreover, the group LASSO yields a larger bias than the concave 2-norm group selection.
\end{remark}

\section{Integrative analysis under the heterogeneity model}

Under this model, two-level selection is needed and can be achieved using composite penalization and sparse group penalization. Properties of composite penalization have been studied in single-dataset analysis, however, under much simpler data and model settings. For sparse group penalization built on concave penalties, properties have not been established for single-dataset analysis.

Define the oracle estimator $\check{\bbeta}=\{\check{\bbeta}_\mathcal {A},0\}$ where
\bqe\label{heteroracle}
\check{\bbeta}_\mathcal {A} ={\arg\min}_{\bbeta_\mathcal {A}}\left\{\frac{1}{2n}\sum_{m=1}^M (\boldY^m- \boldX_{S_m}^m\bbeta_{S_m}^m)^\top W_m (\boldY^m- \boldX_{S_m}^m\bbeta_{S_m}^m)\right\}.
\eqe
Define $\bar{\rho}_{2}^{* m} = \lambda_{\max}\{n_m^{-1}\boldX_{S_m}^{m\top}{ W_m} ^2 \boldX^m_{S_m}\}$, $\underline{\rho}_{1}^{* m}  = \lambda_{\min}\{n_m^{-1}\boldX_{S_m}^{m\top} { W_m} \boldX^m_{S_m}\}$
and \\
$\psi_m^*= \|\boldX_{S_m^c}^{m\top}{ W_m}\boldX^m_{S_m}(\boldX_{S_m}^{m\top} { W_m}\boldX^m_{S_m})^{-1}\|_\infty.$
\begin{theorem}\label{heter:th1}
Consider the estimator defined in (\ref{heteroracle}).
Under Condition 1-3, we have
\[\Pr\left\{\|\check{\bbeta}_{S_m}^{ m} - {\bbeta_{S_m}^{m}}^*\|_2\leq \sqrt{\frac{|S_m|}{n_m}}C_m, m=1,\cdots,M\right\}\geq 1- \sum\limits_{m = 1}^M \exp\left\{-C_m^2\frac{|S_m|(\underline{\rho}_{1}^{* m})^2}{8\bar{\rho}_{2}^{* m}\sigma_m^2 }\right\}\]
with $C_m<\sqrt{\frac{n_m}{|S_m|}}$.
\end{theorem}
\begin{corollary}\label{corollary-3}
Suppose that for $m=1,\cdots, M$, $\underline{\rho}_{1}^{* m}$ and $\bar{\rho}_{2}^{* m} $ are bounded away from zero and infinity, $n/n_m = O(1)$, and $|S|\ll n$. Under Condition 1-3, $\|\check{\bbeta}_{S_m}^{ m} - {\bbeta_{S_m}^{m}}^*\|_2 = O_p(\sqrt{\frac{|S_m|}{n_m}})$ for $m=1,\cdots, M$.
\end{corollary}

\subsection{Composite penalization}

Consider the objective function
\bqe\label{obj4}
L(\bbeta)=\frac{1}{2n}\sum_{m=1}^M (\boldY^m- \boldX^m\bbeta^m)^\top W_m (\boldY^m- \boldX^m\bbeta^m)+\sum_{j=1}^p p_{O,\lambda_O}\left(\sum_{m=1}^M p_{I,\lambda_I}(|\beta_j^m|)\right),
\eqe
where the outer penalty $p_{O,\lambda_O}(\cdot)$ determines the overall importance of a variable, and the inner penalty $p_{I,\lambda_I}(\cdot)$ determines its individual importance. $\lambda_O$ and $\lambda_I$ are tuning parameters. A specific example is the composite MCP (cMCP) where both $p_{O,\lambda_O}$ and $p_{I,\lambda_I}$ are MCP.

\noindent
[{\sc Condition 4}] $\theta_O=\inf\left\{\frac{|t|}{\lambda_O}:~\frac{p'_{O,\lambda_O}(|t|)}{\lambda_O} = 0\right\}$ and
$\theta_I = \inf\left\{\frac{|t|}{\lambda_I}: ~\frac{p'_{I,\lambda_I}(|t|)}{ \lambda_I}=0\right\}$ are bounded.

Denote $J^{-m}=\max\left\{\sum\limits_{i\neq m}^M I(\beta_j^i\neq 0),j\in S-S_m\right\}$ and $f_I^{\max}= \max_t  p_{I,\lambda_I}(t)$.

\begin{theorem}\label{heter:th2}
Consider the minimizer of (\ref{obj4}). Assume Condition 1-2 and 4. Set $$C_m^\dag \leq \frac{\min\limits_{(j,m)\in \mathcal {A}}|{\beta_j^m}^*|}{2}\sqrt{\frac{n_m}{|S_m|}},~ \lambda_I < \frac{\min\limits_{(j,m)\in \mathcal {A}}|{\beta_j^m}^*|}{2\theta_I}, ~\lambda_O\theta_O>f_I^{\max}\max_m{(J^{-m})}.$$ Then $\check{\bbeta}$ is a local minimizer with probability at least $1-\tau_2$, where
\bse
\tau_2&=& \sum\limits_{m = 1}^M \exp\left\{-C_m^{\dag 2}\frac{|S_m|(\underline{\rho}_{1}^{* m})^2}{8\bar{\rho}_{2}^{* m}\sigma_m^2 }\right\}+2|S| \sum\limits_{m=1}^M\exp\left\{-\frac{n^2 p'^2_{I,\lambda_I}(0+)p'^2_{O,\lambda_O}(J^{-m} f_I^{max})}{2n_m \bar{\rho}_{2}^{* m}\sigma_m^2(1+\psi^*_m)^2}\right\}\\
&&~~~~+2 (p-|S|) \sum\limits_{m=1}^M\exp\left\{-\frac{n^2 p'^2_{I,\lambda_I}(0+)p'^2_{O,\lambda_O}(0+)}{2n_m \Lambda_m \sigma_m^2(1+\psi^*_m)^2}\right\}.
\ese
\end{theorem}
This theorem establishes the consistency of composite penalized estimates. A simplified statement is provided in the following corollary.
\begin{corollary}\label{corollary-4}
Suppose that for $m=1,\cdots, M$, $\underline{\rho}_{1}^{* m}, \bar{\rho}_{2}^{* m}$, and $\Lambda_m$ are bounded away from zero and infinity, $n/n_m = O(1)$, $|S|\ll n$, $\log p = O(n^\alpha)$ with $\alpha<1$, and $\psi_m^* = O(n^{\alpha_1})$ with $\alpha_1\in[0,1/2)$. Under Condition 1,2 and 4, if $\lambda_I < \frac{\min\limits_{(j,m)\in \mathcal {A}}|{\beta_j^m}^*|}{2\theta_I}$, $\lambda_O\theta_O = Mf_I^{\max}$, and $\lambda_I \lambda_O \gg n^{\frac{\alpha-1}{2}+\alpha_1}$, composite penalization can achieve the two-level selection consistency.
\end{corollary}
\begin{remark}
\cite{liu2012integrative} also suggests the composition of MCP and LASSO. We conjuncture that it is estimation consistent, can consistently identify the overall importance of variables, but in general is not consistent at the individual level.
\end{remark}

\subsection{Sparse group penalization}

Consider the objective function
\bqe\label{obj5}
L(\bbeta)=\frac{1}{2n}\sum_{m=1}^M (\boldY^m- \boldX^m\bbeta^m)^\top W_m (\boldY^m- \boldX^m\bbeta^m)+\sum_{j=1}^p p_{1,\lambda_1}(\|\bbeta_j\|_2)+\sum_{j=1}^p \sum_{m=1}^M p_{2,\lambda_2}(|\beta_j^m|).
\eqe
$\lambda_1$ and $\lambda_2$ are tuning parameters. Here the penalty is the sum of group and individual penalties. The first penalty determines the overall importance of a variable, and the second penalty determines its individual importance.

Consider penalties $p_{1,\lambda_1}$ and $p_{2,\lambda_2}$ that satisfy Condition 2 and 4 with bounded constants $\theta_1$ and $\theta_2$. Consider the estimator defined by minimizing (\ref{obj5}).
\begin{theorem}\label{heter:th3}
Suppose that Condition 1-2 and 4 hold. Set $$C_m^\dag \leq \frac{\min\limits_{(j,m)\in \mathcal {A}}|{\beta_j^m}^*|}{2}\sqrt{\frac{n_m}{|S_m|}},~ \lambda_1 < \frac{\min\limits_{j\in S}\|{\bbeta_j}^*\|_2}{2\theta_1}, ~\lambda_2< \frac{\min\limits_{(j,m)\in \mathcal {A}}|{\beta_j^m}^*|}{2\theta_2}.$$ Then $\check{\bbeta}$ is a local minimizer with probability at least $1-\tau_3$, where
\bse
\tau_3& =&\sum\limits_{m = 1}^M \exp\left\{-C_m^{\dag 2}\frac{|S_m|(\underline{\rho}_{1}^{* m})^2}{8\bar{\rho}_{2}^{* m}\sigma_m^2 }\right\}+2|S| \sum\limits_{m=1}^M\exp\left\{-\frac{n^2 p'^2_{2,\lambda_2}(0+)}{2n_m \bar{\rho}_{2}^{* m}\sigma_m^2(1+\psi^*_m)^2}\right\}\\
&&+2 (p-|S|) \sum\limits_{m=1}^M\exp\left\{-\frac{n^2 [p'_{1,\lambda_1}(0+)+p'_{2,\lambda_2}(0+)]^2}{2n_m \Lambda_m\sigma_m^2(1+\psi^*_m)^2}\right\}.
\ese
\end{theorem}
That is, the sparse group penalization also enjoys the consistency properties. For theoretical purpose, $p_{1,\lambda_1}$ and $p_{2,\lambda_2}$ do not need to take the same form. However using the same $p_{1,\lambda_1}$ and $p_{2,\lambda_2}$ may facilitate computation. We then derive the following asymptotic result.

\begin{corollary}\label{corollary-5}
Suppose that for $m=1,\cdots, M$, $\underline{\rho}_{1}^{* m}, \bar{\rho}_{2}^{* m}$, and $\Lambda_m$ are bounded away from zero and infinity, $n/n_m = O(1)$, $|S|\ll n$, $\log p = O(n^\alpha)$ with $\alpha<1$, and $\psi_m^* = O(n^{\alpha_1})$ with $\alpha_1\in[0,1/2)$. Under Condition 1-2 and 4, if $\lambda_1 < \frac{\min\limits_{j\in S}\|{\bbeta_j}^*\|_2}{2\theta_1}, ~\lambda_2< \frac{\min\limits_{(j,m)\in \mathcal {A}}|{\beta_j^m}^*|}{2\theta_2}$, $\lambda_1 \gg n^{-\frac{1}{2}+\alpha_1}$ and $\lambda_1+\lambda_2 \gg n^{\frac{\alpha-1}{2}+\alpha_1}$, then the sparse group penalization achieves the two-level selection consistency.
\end{corollary}

\section{Numerical study}

\subsection{Computation}

With the weighted LS approach, the loss function (\ref{obj0}) has a least squares form. In single-dataset analysis with a LS loss, multiple computational algorithms have been developed for group penalization, composite penalization, and sparse group penalization \citep{friedman2010, breheny2009penalized, liu2012integrative}. Here we adopt the existing gradient descent algorithms with minor modifications. Convergence properties can be derived following \cite{breheny2011} and references therein. Details are omitted here. The penalization methods involve the tuning parameter $\lambda (\lambda_I, \lambda_O, \lambda_1, \lambda_2)$. The theorems provide results on the asymptotic order. MCP also involves the additional regularization parameter $a$. Following the literature, we consider a small number of values for $a$, in particular including 1.8, 3, 6 and 10. In numerical study, we use 5-fold cross validation for tuning parameter selection.

\subsection{Simulation}
We simulate three datasets, each with 100 subjects. For each subject, we simulate 1,000 covariates. The covariates have a joint normal distribution, with marginal means equal to zero and variances equal to one. Consider two correlation structures. The first is the auto-regressive (AR) correlation, where covariates $j$ and $k$ have correlation coefficient $\rho^{|j-k|}$. $\rho=0.2$, 0.5, and 0.8, corresponding to weak, moderate, and strong correlations, respectively. The second is the banded correlation. Here three scenarios are considered. Under the first scenario, covariates $j$ and $k$ have correlation coefficient 0.3 if $|j-k|=1$ and 0 otherwise. Under the second scenario, covariates $j$ and $k$ have correlation coefficient 0.6 if $|j-k|=1$, 0.3 if $|j-k|=2$, and 0 otherwise. Under the third scenario, covariates $j$ and $k$ have correlation coefficient 0.6 if $|j-k|=1$, 0.3 if $|j-k|=2$, 0.15 if $|j-k|=3$, and 0 otherwise. Both the homogeneity and heterogeneity models are simulated. Under the homogeneity model, all three datasets share the same twenty important covariates. Under the heterogeneity model, each dataset has twenty important covariates. The three datasets share ten important covariates in common, and the rest important covariates are dataset-specific. Under both models, there are a total of sixty true positives. The nonzero coefficients are randomly generated from a normal distribution with mean zero and variance 0.3125 and 1.25, representing low and high signal levels. The log event times are generated from the AFT models with intercept equal to 0.5 and N(0,1) random errors. The log censoring times are independently generated from uniform distributions. The overall censoring rate is about 30\%.

The simulated data are analyzed using group MCP (GMCP), composite MCP (cMCP), and sparse group MCP (SGMCP). In addition, we also consider two alternatives. The first is a meta-analysis method, where each dataset is analyzed separately using MCP, and then the analysis results are combined across datasets. The second is a pooled analysis method, where the three datasets are combined into a big data matrix, and then variable selection is conducted using MCP. Note that the differences across simulated datasets are smaller than those encountered in practice, which favors meta- and pooled analysis. We acknowledge that multiple other methods are applicable to the simulated data. The two alternatives have the closest framework as the proposed methods.

Summary results based on 200 replicates are shown in Table 1 and 2. Performance of the integrative analysis methods as well as alternatives depend on the similarity of sparsity structures across datasets, correlation structure, and signal level. As an example of the homogeneity model, consider the correlation structure ``Banded 2" in Table 1. The homogeneity model favors GMCP, which identifies 34.7 true positives with an average model size 45.2. The cMCP method identifies fewer true positives (30.5). A large number of false positives are identified, with an average model size 149.7. SGMCP identifies 25.6 true positives, with a very small number of false positives (average model size 27.4). In comparison, the meta-analysis and pooled analysis identify much fewer true positives (17.6 and 16.1, respectively). As an example of the heterogeneity model, consider the correlation structure ``AR $\rho=0.5$" in Table 2. The cMCP method identifies the most true positives (42.1 on average), but at the price of a large number of false positives (average model size 185.1). GMCP identifies 34.6 true positives. However by forcing the same sparsity structure across datasets, it also identifies a considerable number of false positives (average model size 61.0). SGMCP identifies 26.9 true positives with an average model size 30.2. The meta-analysis and pooled analysis methods identify fewer true positives.

\subsection{Analysis of lung cancer prognosis data}

In the U.S., lung cancer is the most common cause of cancer death for both men and women. To identify genetic markers associated with the prognosis of lung cancer, gene profiling studies have been extensively conducted. We follow \cite{XXC} and collect data from four independent studies with gene expression measurements. The UM (University of Michigan Cancer Center) dataset has a total of 92 patients, with 48 deaths during follow-up. The median follow-up is 55 months. The HLM (Moffitt Cancer Center) dataset has a total of 79 patients, with 60 deaths during follow-up. The median follow-up is 39 months. The DFCI (Dana-Farber Cancer Institute) dataset has a total of 78 patients, with 35 deaths during follow-up. The median follow-up is 51 months.  The MSKCC dataset has a total of 102 patients, with 38 deaths during follow-up. The median follow-up is 43.5 months.

Gene expressions were measured using Affymetrix U122 plus 2.0 arrays. A total of 22,283 probe sets were profiled in all four datasets. We first conduct gene expression normalization for each dataset separately, and then normalization across datasets is also conducted to enhance comparability. To further remove noises and improve stability, we conduct a marginal screening and keep the top 2,000 genes for downstream analysis. The expression of each gene in each dataset is normalized to have zero mean and unit variance.

We analyze data using cMCP (Table 3), SGMCP (Table S2.1), meta-analysis (Table S2.2), pooled analysis (Table S2.3), and GMCP (Table S2.4). Although there is overlap, different methods identify significantly different sets of genes. The cMCP method identifies more genes, particularly many more than SGMCP. Such a result fits the pattern observed in simulation. Unlike in simulation, we are not able to objectively evaluate the marker selection results. To provide further insights, we evaluate prediction performance using a cross-validation based approach. Specifically, we split the samples into a training and a testing set with size 3:1. Estimates are generated using the training set samples and used to make prediction for the testing set samples. We separate the testing set samples into two sets with equal sizes based on  $\boldX^m\bbeta^m$'s. The logrank statistic is computed, evaluating survival difference of the two sets. To reduce the risk of an extreme split, we repeat this process 100 times and compute the average logrank statistics as 7.65 (cMCP), 4.95 (SGMCP), 5.35 (meta-analysis), 5.2 (pooled analysis), and 6.45 (GMCP). All methods are able to separate samples into sets with different survival risk. The cMCP method has the best prediction performance (p-value 0.0057).

\section{Discussion}

In this article, we have studied the integrative analysis of survival data under the AFT model. The existing research on this topic has been scattered, and this study is the first to systematically study this complicated problem. Both the homogeneity and heterogeneity models have been considered, along with multiple penalization methods. Significantly advancing from the existing studies, the present study rigorously establishes the selection and estimation consistency properties. Although some theoretical development has been motivated by the existing studies, the heterogeneity across multiple datasets and specific data and model settings make this study unique. Especially, the properties of sparse group penalization have not been studied in single-dataset analysis. Thus this study has both methodological and theoretical contributions. The computational aspect is similar to that in the literature and is largely omitted. Tuning parameter selection using cross validation shows reasonable performance in simulation and data analysis. Theoretical investigation on the consistency of cross validation is very much challenging and postponed.
Another contribution is that this study directly compares different methods. The advantage of GMCP under the homogeneity model is expected. Under the heterogeneity model, cMCP may identify a few more true positives, however, at the price of a large number of false positives. The theoretical study does not provide an explanation to this observation. More studies on finite sample properties are needed. In simulation, a total of 24 settings are considered and show similar patterns. More extensive simulations may be pursued in the future. In data analysis, different methods identify different sets of genes. The observed patterns are similar to those in simulation. In addition, cMCP identifies the most genes but also has the best prediction performance. More extensive, especially biological studies may be needed to fully comprehend the data analysis results. In this study, we have focused on survival data and the AFT model. Extensions to other data and model are of interest to future study.



\clearpage

\begin{table}
\caption{Simulation at the low signal level. In each cell, the first row is the number of true positives (sd), and the second row is the number of model size (sd).}
\centering
{
\begin{tabular}{cccccc}
\hline
Correlation & Meta & Pooled & GMCP & cMCP & SGMCP  \\
\hline
&\multicolumn{5}{c}{\underline{Homogeneity model}}\\
\multirow{2}{*}{AR $\rho=0.2$}
&30.3(5.7)&29.0(8.4)&48.8(6.2)&42.6(4.2)&36.5(6.7)\\
&62.4(19.1)&56.5(29.3)&57.4(9.4)&193.2(13.9)&39.1(8.3)\\
\hline
\multirow{2}{*}{AR $\rho=0.5$}
&20.4(6.0)&18.3(6.7)&39.5(7.9)&33.3(8.1)&28.6(6.9)\\
&38.7(17.7)&31.2(16.3)&50.8(12.4)&160.6(83.0)&30.9(9.1)\\
\hline
\multirow{2}{*}{AR $\rho=0.8$}
&10.9(2.6)&10.3(3.3)&24.8(7.7)&18.3(4.1)&16.8(5.2)\\
&17.9(6.1)&15.5(6.4)&34.4(12.8)&75.4(59.2)&18.6(7.2)\\
\hline
\multirow{2}{*}{Banded 1}
&26.7(5.8)&25.1(7.6)&46.2(7.6)&40.3(4.5)&34.7(6.2)\\
&54.3(18.7)&48.7(26.1)&56.5(12.6)&196.6(12.7)&37.8(8.9)\\
\hline
\multirow{2}{*}{Banded 2}
&17.6(4.5)&16.1(5.0)&34.7(8.3)&30.5(6.0)&25.6(5.9)\\
&30.4(11.6)&25.4(12.5)&45.2(13.7)&149.7(95.0)&27.4(7.2)\\
\hline
\multirow{2}{*}{Banded 3}
&17.7(5.3)&16.2(4.9)&37.3(7.3)&31.4(5.8)&26.1(6.3)\\
&32.1(18.6)&26.8(12.9)&51.1(13.7)&166.3(81.7)&28.2(7.6)\\
\hline
&\multicolumn{5}{c}{\underline{Heterogeneity model}}\\
\multirow{2}{*}{AR $\rho=0.2$}
&21.3(5.1)&20.2(5.7)&26.0(9.0)&37.6(5.2)&22.5(7.2)\\
&35.5(13.8)&31.4(13.9)&53.0(20.3)&199.2(40.3)&28.4(11.0)\\
\hline
\multirow{2}{*}{AR $\rho=0.5$}
&16.8(5.1)&16.7(5.3)&22.8(6.2)&31.7(6.9)&18.8(5.7)\\
&28.5(10.8)&27.3(12.0)&45.5(15.2)&154.8(94.4)&21.9(7.7)\\
\hline
\multirow{2}{*}{AR $\rho=0.8$}
&10.6(3.8)&10.3(3.5)&15.2(5.5)&20.0(4.9)&11.9(4.2)\\
&17.0(6.3)&15.3(6.3)&31.4(12.9)&99.9(84.4)&15.3(6.8)\\
\hline
\multirow{2}{*}{Banded 1}
&20.4(4.8)&19.9(6.0)&25.2(6.7)&35.3(6.7)&20.9(6.0)\\
&35.2(15.2)&31.3(13.9)&48.9(14.5)&172.2(77.9)&24.9(7.9)\\
\hline
\multirow{2}{*}{Banded 2}
&16.1(4.0)&15.1(3.9)&21.4(6.1)&28.0(5.4)&17.5(4.8)\\
&24.9(8.4)&22.8(7.7)&44.0(12.2)&129.9(103.4)&21.0(6.2)\\
\hline
\multirow{2}{*}{Banded 3}
&15.9(3.6)&15.2(4.4)&20.2(6.0)&27.1(6.2)&17.8(4.9)\\
&26.8(10.8)&24.3(10.2)&43.3(14.2)&102.7(115.7)&22.3(7.5)\\
\hline
\end{tabular}
}
\end{table}

\begin{table}
\caption{Simulation at the high signal level. In each cell, the first row is the number of true positives (sd), and the second row is the number of model size (sd).}
\centering
{
\begin{tabular}{cccccc}
\hline
Correlation & Meta & Pooled & GMCP & cMCP & SGMCP  \\
\hline
&\multicolumn{5}{c}{\underline{Homogeneity model}}\\
\multirow{2}{*}{AR $\rho=0.2$}
&39.4(4.5)&39.2(5.4)&58.3(2.3)&52.3(2.9)&49.9(3.8)\\
&49.9(9.3)&48.8(11.4)&60.1(4.2)&174.6(11.6)&50.1(4.1)\\\hline
\multirow{2}{*}{AR $\rho= 0.5 $}
&30.1(5.0)&30.0(6.0)&55.4(3.6)&46.5(3.3)&44.2(4.0)\\
&42.0(10.1)&41.8(12.3)&58.3(4.3)&179.8(15.3)&44.5(4.2)\\\hline
\multirow{2}{*}{AR $\rho= 0.8 $}
&17.4(3.8)&17.1(3.9)&46.5(6.7)&29.5(6.4)&29.6(5.9)\\
&24.2(6.5)&23.6(7.3)&54.1(10.6)&103.8(97.8)&30.7(6.1)\\\hline
\multirow{2}{*}{Banded  1 }
&36.9(4.7)&35.9(5.1)&57.2(2.7)&50.3(2.9)&47.9(4.3)\\
&47.3(8.4)&43.7(7.6)&58.7(4.4)&178.4(12.1)&48.3(4.4)\\\hline
\multirow{2}{*}{Banded  2 }
&25.9(4.3)&25.5(4.7)&53.3(4.6)&41.1(3.4)&38.6(5.5)\\
&36.3(8.8)&34.4(9.1)&57.8(8.3)&186.2(16.7)&39.7(6.1)\\\hline
\multirow{2}{*}{Banded  3 }
&27.1(3.8)&26.5(4.3)&53.7(4.5)&42.4(4.4)&40.8(4.7)\\
&37.3(8.4)&35.8(8.3)&57.8(7.0)&179.8(21.4)&42.0(5.6)\\
\hline
&\multicolumn{5}{c}{\underline{Heterogeneity model}}\\
\multirow{2}{*}{AR $\rho= 0.2 $}
&34.4(4.1)&34.0(4.1)&40.0(4.2)&48.91(3.2)&33.9(4.6)\\
&39.7(6.0)&37.9(4.8)&69.2(7.9)&180.4(18.9)&36.6(4.7)\\\hline
\multirow{2}{*}{AR $\rho= 0.5 $}
&25.9(4.5)&24.1(5.9)&34.6(5.7)&42.1(4.1)&26.9(4.8)\\
&32.7(6.6)&29.5(7.3)&61.0(9.8)&185.1(18.0)&30.2(6.2)\\\hline
\multirow{2}{*}{AR $\rho= 0.8 $}
&16.4(3.4)&15.6(3.5)&23.7(5.6)&26.8(5.3)&17.5(4.4)\\
&22.2(5.1)&21.3(6.5)&44.3(10.3)&157.5(87.3)&20.9(5.6)\\\hline
\multirow{2}{*}{Banded  1 }
&30.8(4.1)&30.2(4.6)&36.8(5.3)&45.8(3.1)&30.0(5.2)\\
&36.0(5.8)&35.4(6.7)&64.1(9.3)&177.7(17.3)&32.6(6.7)\\\hline
\multirow{2}{*}{Banded  2 }
&22.9(4.6)&22.4(4.1)&32.1(5.9)&36.6(4.3)&25.2(4.9)\\
&29.3(7.8)&27.5(5.4)&57.4(8.4)&169.2(51.2)&28.6(5.3)\\\hline
\multirow{2}{*}{Banded  3 }
&23.0(4.6)&22.6(4.2)&31.6(6.2)&37.4(5.0)&24.2(6.8)\\
&28.7(5.8)&27.9(5.3)&57.1(9.9)&169.2(42.1)&26.6(7.5)\\
\hline
\end{tabular}
}
\end{table}

\begin{table}
\caption{Analysis of lung cancer data using cMCP: identified genes and their estimates.}
\centering
{
\begin{tabular}{cccccc}
\hline
Probe & Gene &  UM & HLM & DFCI & MSKCC  \\
\hline
201462$\_$at	&	SCRN1	&		&	0.0045	&		&		\\
202637$\_$s$\_$at	&	ICAM1	&		&		&		&	0.0037	\\
203240$\_$at	&	FCGBP	&		&	0.0024	&		&		\\
203876$\_$s$\_$at	&	MMP11	&		&		&		&	-0.0013	\\
203917$\_$at	&	CXADR	&		&		&		&	0.0040	\\
203921$\_$at	&	CHST2	&	0.0024	&		&		&		\\
204855$\_$at	&	SERPINB5	&	-0.0008	&		&		&		\\
205234$\_$at	&	SLC16A4	&		&		&		&	-0.0016	\\
205399$\_$at	&	DCLK1	&		&		&		&	-0.0031	\\
206461$\_$x$\_$at	&	MT1H	&		&		&	-0.0008	&		\\
206754$\_$s$\_$at	&	CYP2B6	&		&		&	0.0048	&		\\
206994$\_$at	&	CST4	&		&	-0.0017	&		&		\\
207850$\_$at	&	CXCL3	&		&	-0.0155	&		&		\\
208025$\_$s$\_$at	&	HMGA2	&		&		&	-0.0016	&		\\
208451$\_$s$\_$at	&	C4A	&		&	0.0038	&		&		\\
208607$\_$s$\_$at	&	SAA2	&		&		&		&	0.0044	\\
209343$\_$at	&	EFHD1	&		&		&		&	0.0028	\\
212328$\_$at	&	LIMCH1	&		&		&	0.0028	&		\\
212338$\_$at	&	MYO1D	&		&	0.0019	&		&		\\
213338$\_$at	&	TMEM158	&		&		&	-0.0003	&		\\
214452$\_$at	&	BCAT1	&		&		&		&	0.0004	\\
215867$\_$x$\_$at	&	CA12	&		&	-0.0054	&		&		\\
218677$\_$at	&	S100A14	&		&		&		&	-0.0081	\\
219654$\_$at	&	PTPLA	&		&	-0.0109	&		&		\\
219747$\_$at	&	NDNF	&	0.0001	&		&		&		\\
220952$\_$s$\_$at	&	PLEKHA5	&		&	-0.0018	&		&		\\
221841$\_$s$\_$at	&	KLF4	&	-0.0024	&		&		&		\\
222043$\_$at	&	CLU	&		&	0.0008	&		&		\\
\hline
\end{tabular}
}
\end{table}

\begin{center}
{\bf\Large  Appendix 
}\\

\end{center}

\setcounter{section}{0}
\setcounter{equation}{0}
\setcounter{table}{0}
\def\theequation{S\arabic{section}.\arabic{equation}}
\def\thesection{S\arabic{section}}
\def\thetable{S\arabic{section}.\arabic{table}}

This file contains proofs (Section S1) for the theoretical results described in the main text as well as additional numerical results (Section S2).

\section{Proofs}

Let
\begin{equation}
\label{notion1}
\boldy^m = {W_m}^{1/2}\boldY^m ~\mbox{ and} ~ X^m = {W_m}^{1/2}\boldX^m.
\end{equation}

Then $(\boldY^m- \boldX^m\bbeta^m)^\top W_m (\boldY^m- \boldX^m\bbeta^m)$
can be rewritten as $\|\boldy^m- X^m\bbeta^m\|^2$, where $\|\cdot\|$ is the $\ell_2$ norm. Moreover, we can easily see that
\begin{equation}\label{notion2}
\boldy^m = X^m \bbeta^m + {W_m}^{1/2}\boldepsilon^m.
\end{equation}

\noindent{\bf Proof of Theorem 1.} ~First, we prove that
\[\Pr\left\{\|\hat{\bbeta}_S^m - {\bbeta_{S}^m}^*\|_2< \lambda \frac{4}{\underline{\rho}_{1}^m }\frac{n}{n_m}, m=1,\cdots,M\right\}\geq 1-\tau_1,\]
where $\tau_1= \sum\limits_{m=1}^M \exp\left(-\frac{\lambda^2 n^2}{2 \sigma_m^2 \bar{\rho}_{2}^m n_m }\right)$.
Recall that $\hat{\bbeta}_\mathcal {B}={\arg\min}_{\bbeta_\mathcal {B}}L(\bbeta_\mathcal {B})$, where
\bse
L(\bbeta_\mathcal {B})=\frac{1}{2n}\sum_{m=1}^M \|\boldy^m- X_S^m\bbeta_S^m\|^2+\lambda \sum_{j\in S} \|\bbeta_j\|_2.
\ese
Let $r_m= \lambda \sqrt{|S|} \frac{4}{\underline{\rho}_{1}^m }\frac{n}{n_m}$ and $\mathfrak{I}=\left\{\bbeta_\mathcal {B}: \|\bbeta_S^m - {\bbeta_{S}^m}^*\|_2 = r_m, m=1,\cdots,M\right\}$.
It suffices to show that
\bse
\Pr\Big(\inf\limits_{
\bbeta_\mathcal {B}\in \mathfrak{I}}L(\bbeta_\mathcal {B})>L(\bbeta_\mathcal {B}^*)\Big)\geq 1-\tau_1.
\ese This implies that with probability at least $1-\tau_1$, $L(\bbeta_\mathcal {B})$ has a local
minimum $\hat{\bbeta}_\mathcal {B}$ that satisfies
$\|\hat{\bbeta}_S^m - {\bbeta_{S}^m}^*\|_2< \lambda \sqrt{|S|} \frac{4}{\underline{\rho}_{1}^m }\frac{n}{n_m}$, for $m=1,\cdots,M$.

Let $\boldu\in R^{p\times M}$ with $\|\boldu_S^m\|_2=1,~ m=1,\cdots,M$. Define $\bbeta_S^m={\bbeta_{S}^m}^* + r_m \boldu_S^m$.
Consider  $Q({\boldu_\mathcal {B}})=n\left\{L(\bbeta_\mathcal {B})-L(\bbeta_\mathcal {B}^*)\right\}$. Obviously, it is equivalent to
show that
\bqe\label{eq:sup}
\Pr\Big(\inf\limits_{
\|\boldu^m\|_2=1,~m=1,\cdots,M}Q({\boldu_\mathcal {B}})>0\Big)\geq 1-\tau_1.
\eqe
Together with (\ref{notion1}) and (\ref{notion2}), we have
\bqe\label{q0}
Q({\boldu_\mathcal {B}}) &=& \frac{1}{2}\sum\limits_{m = 1}^M {\left( {{{\left\| {{\boldy^m} - {X_S^m}\left( {{\bbeta _{S}^m}^* + {r_m}\boldu_S^m} \right)} \right\|}^2} - {{\left\| {{\boldy^m} - {X_S^m}{\bbeta _{S}^m}^*} \right\|}^2}} \right)} \nonumber\\
 &&+ n\lambda \sum\limits_{j \in S} {\left\{ { {{{\left\| {{\bbeta _{j}^*} + \boldr \circ {\boldu_j}} \right\|}_2}}  -  {{{\left\| {{\bbeta _{j}^*}} \right\|}_2}}} \right\}}\nonumber\\
 &=&  - \sum\limits_{m = 1}^M {{r_m}\boldu{{_S^m}^{\rm \top}}{\boldX_S^m}^{\rm \top}{W_m}{\boldepsilon ^m}}  + \frac{1}{{2}}\sum\limits_{m = 1}^M {r_m^2\boldu{{_S^m}^{\rm \top}}{\boldX_S^m}^{\rm \top}{W_m}{\boldX_S^m}\boldu_S^m}  \nonumber\\
 &&+  n\lambda \sum\limits_{j \in S} {\left\{ { {{{\left\| {{\bbeta _{j}^*} + \boldr \circ {\boldu_j}} \right\|}_2}}  -  {{{\left\| {{\bbeta _{j}^*}} \right\|}_2}}} \right\}} \nonumber\\
&=:& Q_1 + Q_2 + Q_3,
\eqe
where $\boldr =(r_1,\cdots,r_M)^\top$, and $\circ$ denotes the Hadamard (component-wise) product.
Write $ Q_1= \sum\limits_{m = 1}^M Q_{1m}$ where $Q_{1m}= -{r_m}\boldu{{_S^m}^{\rm \top}}{\boldX_S^m}^{\rm \top}{W_m}{\boldepsilon ^m}$.
Note that $\|W_m\boldX_S^m\boldu{{_S^m}}\|_2^2\leq  n_m\bar{\rho}_{2}^m $. With the sub-Gaussian tail as specified in Condition 1, we have for any given $\varepsilon_m$
\bse
\Pr(|Q_{1m}|>r_m \varepsilon_m)\leq 2\exp\left(-\frac{\varepsilon_m^2}{2\sigma_m^2 \|W_m\boldX_S^m\boldu{{_S^m}}\|_2^2}\right)
\leq 2\exp\left(-\frac{\varepsilon_m^2}{2 n_m \bar{\rho}_{2}^m  \sigma_m^2 }\right).
\ese
Together with the Bonferroni's inequality, we have
\bse
\Pr(Q_1 <-\sum\limits_{m = 1}^M r_m  \varepsilon_m) \leq \sum_{m=1}^M \Pr(Q_{1m}<-r_m \varepsilon_m) \leq \sum_{m=1}^M \exp\left(-\frac{\varepsilon_m^2}{2 n_m \bar{\rho}_{2}^m  \sigma_m^2 }\right).
\ese
Set $ \varepsilon_m = \frac{1}{4} \underline{\rho}_{1}^m  n_m r_m $. Then
\bqe \label{q1}
\Pr(Q_1\geq -\frac{1}{4}\sum\limits_{m = 1}^M r_m^2 n_m\underline{\rho}_{1}^m)  \geq 1- \sum_{m=1}^M \exp\left(-\frac{n_m r_m^2(\underline{\rho}_{1}^m )^2}{32  \bar{\rho}_{2}^m \sigma_m^2 }\right).
\eqe
For $Q_2$, since $\boldu{{_S^m}^{\rm \top}}{\boldX_S^m}^{\rm \top}W_m{\boldX_S^m}\boldu_S^m \geq n_m\underline{\rho}_{1}^m $, we have
\bqe \label{q2}
Q_2 \geq \frac{ 1 }{{2}}\sum\limits_{m = 1}^M {r_m^2 n_m\underline{\rho}_{1}^m  }.
\eqe
Term $Q_3$ can be dealt with as follows.
By the Triangle inequality and $(\sum\limits_{i=1}^d |v_i|)^2 \leq d\sum\limits_{i=1}^d v_i^2$, for any sequence ${v_i}$, we have
\bse
&&\sum\limits_{j \in S}{\left\| {{\bbeta _{j}^*} + \boldr \circ {\boldu_j}} \right\|}_2-{\left\| {{\bbeta _{j}^*}} \right\|}_2 \leq \sum\limits_{j \in S}\left\|\boldr \circ {\boldu_j}\right\|_2\\
&& ~~~~~\leq \sqrt{|S|}\sqrt{\sum\limits_{j \in S}\left\|\boldr \circ {\boldu_j}\right\|_2^2}=\sqrt{|S|}\sqrt{\sum\limits_{m = 1}^M {r_m^2}}
 \leq \sqrt{|S|}\sum\limits_{m = 1}^M {r_m}.
\ese
Therefore, we have that term $Q_3$ satisfies
\bqe\label{q3}
|Q_{3}| \leq n\lambda \sqrt{|S|}\sum\limits_{m = 1}^M {r_m}.
\eqe
Combining (\ref{q0}), (\ref{q1}), (\ref{q2}), and (\ref{q3}), we have
\bqe
Q({\boldu_S})\geq \frac{ 1}{{4}}\sum\limits_{m = 1}^M {r_m^2 n_m\underline{\rho}_{1}^m  }- n\lambda \sqrt{|S|}\sum\limits_{m = 1}^M {r_m}:=L(\boldr)
\eqe
with probability at least
$1-\sum\limits_{m=1}^M \exp\left(-\frac{ n_m r_m^2(\underline{\rho}_{1}^m )^2}{32  \bar{\rho}_{2}^m  \sigma_m^2}\right)$. Recall that $r_m=\lambda \sqrt{|S|} \frac{4}{\underline{\rho}_{1}^m }\frac{n}{n_m}.$
Then $L(\boldr)>0$ with probability at least
$
1-\sum\limits_{m=1}^M \exp\left(-\frac{\lambda^2 |S| n^2}{2  \sigma_m^2\bar{\rho}_{2}^m n_m  }\right).
$
Therefore, (\ref{eq:sup}) is proved, and Part 1 of Theorem 1 is established.

Now consider Part 2. By the Karush-Kuhn-Tucher(KKT) conditions, we need to prove that for $m =1,\cdots,M$,
\bqe
&&-X_S^{m \rm \top}\left(\boldy^m-X_{S}^{m}\hat{\bbeta}_{S}^m\right)
        +n\lambda \frac{\hat{\bbeta}_{S}^m}{\|\hat{\bbeta}_\mathcal {B}\|_2}=0,\label{kkt1:gl}\\
&&\|X_{S^c}^{\rm \top}(\boldy^m- X_S^m\hat{\bbeta}_S^m)\|_\infty\leq n\lambda. \label{kkt2:gl}
\eqe
Then $\hat{\bbeta}^{glasso}=\{\hat{\bbeta}^{glasso}_\mathcal {B}, \hat{\bbeta}^{glasso}_{\mathcal {B}^c}\}$ with $\hat{\bbeta}^{glasso}_\mathcal {B}=\hat{\bbeta}_\mathcal {B}, \hat{\bbeta}^{glasso}_{\mathcal {B}^c}=0$ is a local minimizer of (3).
From Part 1, $\tilde{\bbeta}_S$ minimizes
\bse
L(\bbeta_\mathcal {B})=\frac{1}{2n}\sum_{m=1}^M \|\boldy^m- X_S^m\bbeta_S^m\|^2+\lambda\sum_{j\in S} \|\bbeta_j\|_2.
\ese
Therefore, (\ref{kkt1:gl}) holds, together with (\ref{notion2}) which also yields
\bqe\label{t1}
\hat{\bbeta}_{S}^m-{\bbeta_{S}^{m}}^*=\left(X_{S}^{m \rm \top} X_{S}^{m}\right)^{-1}\left\{X_{S}^{m \rm \top}{W_m}^{1/2}\boldepsilon^m-n\lambda \frac{\hat{\bbeta}_{S}^m}{\|\hat{\bbeta}_{\mathcal {B}}\|_2}\right\}.
\eqe
Note that
\bqe\label{t2}
X_{S^c}^{m \rm \top}(\boldy^m- X_S^m\hat{\bbeta}_S^m)=X_{S^c}^{m \rm \top}{W_m}^{1/2}\boldepsilon^m -X_{S^c}^{m \rm \top}X_S^m(\hat{\bbeta}_S^m-{\bbeta_{S}^m}^*).
\eqe
Substituting (\ref{t1}) into (\ref{t2}), we obtain
\bqe\label{tnorm}
&&\|X_{S^c}^{m \rm \top}(\boldy^m- X_S^m\hat{\bbeta}_S^m)\|_\infty \nonumber\\
&=&\left\|X_{S^c}^{m \rm \top}{W_m}^{1/2}\boldepsilon^m -X_{S^c}^{m \rm \top}X_S^m\left(X_{S}^{m \rm \top} X_{S}^{m}\right)^{-1}\left\{X_{S}^{m \rm \top}{W_m}^{1/2}\boldepsilon^m-n\lambda  \frac{\hat{\bbeta}_{S}^m}{\|\hat{\bbeta}_{\mathcal {B}}\|_2}\right\}\right\|_\infty\nonumber\\
&\leq& \left\|X_{S^c}^{m \rm \top}{W_m}^{1/2}\boldepsilon^m\right\|_\infty + \left\|X_{S^c}^{m \rm \top}X_S^m\left(X_{S}^{m \rm \top} X_{S}^{m}\right)^{-1}X_{S}^{m \rm \top}{W_m}^{1/2}\boldepsilon^m\right\|_\infty \nonumber\\
&&+ n\lambda\left\|X_{S^c}^{m \rm \top}X_S^m\left(X_{S}^{m \rm \top} X_{S}^{m}\right)^{-1} \frac{\hat{\bbeta}_{S}^m}{\|\hat{\bbeta}_{\mathcal {B}}\|_2}\right\|_\infty\nonumber\\
&\leq& \left\|\boldX_{S^c}^{m \rm \top}W_m\boldepsilon^m\right\|_\infty + \left\|\boldX_{S^c}^{m \rm \top}{W_m}\boldX_S^m\left(\boldX_{S}^{m \rm \top}{W_m} \boldX_{S}^{m}\right)^{-1}\right\|_\infty \left\|\boldX_{S}^{m \rm \top}{W_m}\boldepsilon^m\right\|_\infty \nonumber\\
&&+ n\lambda\left\|\boldX_{S^c}^{m \rm \top}{W_m}\boldX_S^m\left(\boldX_{S}^{m \rm \top}{W_m} \boldX_{S}^{m}\right)^{-1}\right\|_\infty \left\|\frac{\hat{\bbeta}_{S}^m}{\|\hat{\bbeta}_{\mathcal {B}}\|_2}\right\|_\infty\nonumber\\
&\leq& \left\|\boldX_{S^c}^{m \rm \top}W_m\boldepsilon^m\right\|_\infty + \psi_m\left\|\boldX_{S}^{m \rm \top}W_m\boldepsilon^m\right\|_\infty +  n\lambda \psi_m
\eqe
By the condition $\psi_m \leq D_m<1$, if
\bqe\label{pb}
\left\|\boldX^{m \rm \top}W_m \boldepsilon^m\right\|_\infty \leq n\lambda \frac{1-D_m}{1+D_m},
\eqe
then from (\ref{tnorm}) it follows
\bse
\|X_{S^c}^{m \rm \top}(\boldy^m- X_S^m\tilde{\bbeta}_S^m)\|_\infty &\leq& \left\|\boldX^{m \rm \top}W_m\boldepsilon^m\right\|_\infty(1+\psi_m)+ n\lambda \psi_m \\
&\leq& n\lambda (1-D_m) + n\lambda D_m=n\lambda.
\ese
We now derive the probability bounds for the event in (\ref{pb}). By the Bonferroni's inequality and sub-Gaussian tail probability bound in Condition 1,
\bqe\label{tail1}
&&\Pr\left\{\left\|\boldX^{m \rm \top}W_m\boldepsilon^m\right\|_\infty > n\lambda  \frac{1-D_m}{1+D_m},~ \mbox{for}~m=1,\cdots,M\right\}\nonumber\\
&\leq& p\sum\limits_{m=1}^M\Pr\left\{|\boldX_j^{m \rm \top}W_m\boldepsilon^m| > n\lambda  \frac{1-D_m}{1+D_m}\right\}\nonumber\\
&\leq& 2p \sum\limits_{m=1}^M\exp\left\{-\frac{n^2\lambda^2 (1-D_m)^2}{2n_m \Lambda_m \sigma_m^2(1+D_m)^2}\right\}.
\eqe
Then  Part 2 is established by combining Part1, (\ref{kkt1:gl}), (\ref{kkt2:gl}), and (\ref{tail1}).
\hfill $\Box$

\vspace{1cm}

\noindent{\bf Proof of Theorem 2.}~ Recall that $\tilde{\bbeta}_\mathcal {B}={\arg\min}_{\bbeta_\mathcal {B}}H(\bbeta_\mathcal {B})$, where
\bse
H(\bbeta_\mathcal {B})=\frac{1}{2n}\sum_{m=1}^M \|\boldy^m- X_S^m\bbeta_S^m\|^2.
\ese
Let $r_m= \sqrt{\frac{|S|}{n}}R_m $ with $R_m\in (0, \infty)$ and $\mathfrak{I}=\left\{\bbeta_\mathcal {B}: \|\bbeta_S^m - {\bbeta_{S}^m}^*\|_2 = r_m, m=1,\cdots,M\right\}$.
Similar as the proof of part 1 in Theorem 1, if we can prove
\bqe\label{th2-1}
\Pr\Big(\inf\limits_{
\bbeta_\mathcal {B}\in \mathfrak{I}}H(\bbeta_\mathcal {B})>H(\bbeta_\mathcal {B}^*)\Big)\geq 1-\sum\limits_{m = 1}^M \exp\left\{-R_m^2\frac{|S|(\underline{\rho}_{1}^m )^2}{8\bar{\rho}_2^m\sigma_m^2 }\right\},
\eqe then $H(\bbeta_\mathcal {B})$ has a local
minimum $\hat{\bbeta}_\mathcal {B}$ that satisfies
$\|\hat{\bbeta}_S^m - {\bbeta_{S}^m}^*\|_2< r_m, m=1,\cdots,M$ with probability at least $1-\sum\limits_{m = 1}^M \exp\left\{-R_m^2\frac{|S|(\underline{\rho}_{1}^m )^2}{8\bar{\rho}_2^m\sigma_m^2 }\right\}$.

Together with (\ref{notion1}) and (\ref{notion2}), we have
\bqe\label{22-1}
H(\bbeta_\mathcal {B})-H(\bbeta_\mathcal {B}^*)
 &=&  - \sum\limits_{m = 1}^M {(\hat{\bbeta}_S^m - {\bbeta_{S}^m}^*)^\top{\boldX_S^m}^{\rm \top}{W_m}{\boldepsilon ^m}}  \nonumber\\
  && + \frac{1}{{2}}\sum\limits_{m = 1}^M {(\hat{\bbeta}_S^m - {\bbeta_{S}^m}^*)^\top{\boldX_S^m}^{\rm \top}{W_m}{\boldX_S^m}(\hat{\bbeta}_S^m - {\bbeta_{S}^m}^*)}  \nonumber\\
&=:& H_1 + H_2,
\eqe
For $H_2$, since $\lambda_{\min}\left\{n_m^{-1}{\boldX_S^m}^{\rm \top}W_m{\boldX_S^m}\right\}=\underline{\rho}_{1}^m $ and $\|\bbeta_S^m - {\bbeta_{S}^m}^*\|_2 = r_m$, we have
\bqe \label{22-2}
H_2 \geq \frac{ 1 }{{2}}\sum\limits_{m = 1}^M {r_m^2 n_m\underline{\rho}_{1}^m  }.
\eqe
For $H_1$ we have for any $\varepsilon_m$,
\bse
\Pr(H_1 \leq-\sum\limits_{m = 1}^M r_m  \varepsilon_m) &\leq& \sum_{m=1}^M\exp\left(-\frac{r_m^2 \varepsilon_m^2}{2\sigma_m^2 \|W_m\boldX_S^m (\hat{\bbeta}_S^m - {\bbeta_{S}^m}^*)\|_2^2}\right) \nonumber\\
&\leq& \sum_{m=1}^M \exp\left(-\frac{\varepsilon_m^2}{2 n_m \bar{\rho}_{2}^m  \sigma_m^2 }\right).
\ese
The first inequality holds due to the sub-Gaussian tail probability under Condition 1, and the last inequality holds due to the fact that $\|W_m\boldX_S^m (\hat{\bbeta}_S^m - {\bbeta_{S}^m}^*)\|_2^2\leq n_m \bar{\rho}_2^m r_m^2 $. Set $ \varepsilon_m = \frac{1}{2} \underline{\rho}_{1}^m  n_m r_m $. Then
\bqe \label{22-3}
\Pr(H_1> -\frac{1}{2}\sum\limits_{m = 1}^M r_m^2 n_m\underline{\rho}_{1}^m)  \geq 1- \sum_{m=1}^M \exp\left(-\frac{n_m r_m^2(\underline{\rho}_{1}^m )^2}{8  \bar{\rho}_{2}^m \sigma_m^2 }\right).
\eqe
Recall that $r_m = \sqrt{\frac{|S|}{n}}R_m $. Combining (\ref{22-1}), (\ref{22-2}) and (\ref{22-3}), we have
(\ref{th2-1}) holds.
This complete the proof of Part 1.

Next, we prove Part 2. By the Karush-Kuhn-Tucher(KKT) conditions, we need to prove that $\hat{\bbeta}^{oracle}$ satisfies
\bqe
&&-X_S^{m \rm \top}\left(\boldy^m-X_{S}^{m}\tilde{\bbeta}_{S}^m\right)
        +n p'_\lambda(\|\tilde{\bbeta}_\mathcal {B}\|_2) \circ\frac{\tilde{\bbeta}_{S}^m}{\|\tilde{\bbeta}_\mathcal {B}\|_2}=0,\label{kkt1:homo}\\
&&\|X_{S^c}^{m \rm \top}(\boldy^m- X_S^m\tilde{\bbeta}_S^m)\|_\infty\leq np'_\lambda(0+). \label{kkt2:homo}
\eqe
If $\min\limits_{j \in S} \|\tilde{\bbeta}_j\|_2>\theta\lambda$, $p'_\lambda(\|\tilde{\bbeta}_\mathcal {B}\|_2)=0$, and certainly (\ref{kkt1:homo}) holds. Define $$R_m^\dag \leq \frac{\min\limits_{j\in S}\|\bbeta_j^*\|_2}{2\sqrt{M}}\sqrt{\frac{n_m}{|S|}}.$$
Note that $\lambda < \frac{\min\limits_{j\in S}\|\bbeta_j^*\|_2}{2\theta}$. Therefore, we can conclude the event
$$\left\{\|\tilde{\bbeta}_S^m-{\bbeta_S^m}^*\|_2\leq \sqrt{\frac{|S|}{n_m}}R_m^\dag ,~m=1,\cdots,M\right\}$$ belongs to the event $\left\{\min\limits_{j \in S} \|\tilde{\bbeta}_j\|_2>\theta\lambda\right\}$. That is,
\bqe\label{tt1}
&&\Pr\left\{\min\limits_{j \in S} \|\tilde{\bbeta}_j\|_2>\theta\lambda\right\}\geq \Pr\left(\|\tilde{\bbeta}_S^m-{\bbeta_S^m}^*\|_2\leq  \sqrt{\frac{|S|}{n_m}}R_m^\dag,~ m=1,\cdots,M\right)\nonumber\\
&& ~~~~~ \geq 1- \sum\limits_{m = 1}^M \exp\left\{-\frac{|S|(\underline{\rho}_{1}^m )^2}{8\sigma_m^2 \bar{\rho}_2^m  }R_m^{\dag 2}\right\}.
\eqe
Now consider the probability of
\bqe\label{kkt}
\|X_{S^c}^{m \rm \top}(\boldy^m- X_S^m\tilde{\bbeta}_S^{ m})\|_\infty\leq np'_\lambda(0+), ~\mbox{for}~m =1,\cdots,M.
\eqe
Note that
\bqe\label{ss1}
X_{S^c}^{m \rm \top}(\boldy^m- X_S^m\tilde{\bbeta}_S^{ m})=\boldX_{S^c}^{m \rm \top}W_m\boldepsilon^m -\boldX_{S^c}^{m \rm \top}W_m \boldX_S^m(\tilde{\bbeta}_S^{ m} -{\bbeta_{S}^m}^*).
\eqe
Combining (\ref{kkt}) and (\ref{ss1}), we can obtain
\bqe\label{snorm}
&&\|X_{S^c}^{m \rm \top}(\boldy^m- X_S^m\tilde{\bbeta}_S^m)\|_\infty \nonumber\\ &=&\left\|\boldX_{S^c}^{m \rm \top}W_m\boldepsilon^m -\boldX_{S^c}^{m \rm \top}W_m \boldX_S^m\left(\boldX_{S}^{m \rm \top}W_m \boldX_{S}^{m}\right)^{-1}\boldX_{S}^{m \rm \top}W_m\boldepsilon^m\right\|_\infty\nonumber\\
&\leq& \left\|\boldX_{S^c}^{m \rm \top}W_m\boldepsilon^m\right\|_\infty + \left\|\boldX_{S^c}^{m \rm \top}W_m \boldX_S^m\left(\boldX_{S}^{m \rm \top}W_m \boldX_{S}^{m}\right)^{-1}\boldX_{S}^{m \rm \top}W_m\boldepsilon^m\right\|_\infty \nonumber\\
&\leq& \left\|\boldX_{S^c}^{m \rm \top}W_m\boldepsilon^m\right\|_\infty + \left\|\boldX_{S^c}^{m \rm \top}W_m \boldX_S^m\left(\boldX_{S}^{m \rm \top}W_m \boldX_{S}^{m}\right)^{-1}\right\|_\infty \left\|\boldX_{S}^{m \rm \top}W_m\boldepsilon^m\right\|_\infty \nonumber\\
&\leq& \left\|\boldX_{S^c}^{m \rm \top}W_m\boldepsilon^m\right\|_\infty + \psi_m\left\|\boldX_{S}^{m \rm \top}W_m\boldepsilon^m\right\|_\infty  \nonumber\\
&\leq&
\left\|\boldX^{m \rm \top}W_m\boldepsilon^m\right\|_\infty (1+\psi_m).
\eqe
If
\bqe\label{sm}
\left\|\boldX^{m \rm \top}W_m\boldepsilon^m\right\|_\infty \leq  \frac{np'_\lambda(0+)}{(1+\psi_m)},
\eqe
then from (\ref{snorm}) it follows
\bse
\|X_{S^c}^{m \rm \top}(\boldy^m- X_S^m\tilde{\bbeta}_S^m)\|_\infty &\leq& \frac{np'_\lambda(0+)}{(1+\psi_m)}(1+\psi_m)
\leq np'_\lambda(0+),
\ese
which proves (\ref{kkt2:homo}).
We now derive the probability bounds for the event in (\ref{sm}). In fact, by the Bonferroni's inequality and sub-Gaussian tail probability bound under Condition 1,
\bqe\label{tt2}
&&\Pr\left\{\left\|\boldX^{m \rm \top}W_m\boldepsilon^m\right\|_\infty > \frac{np'_\lambda(0+)}{(1+\psi_m)},~ \exists~m \in \{1,\cdots,M\}\right\}\nonumber\\
&\leq& p\sum\limits_{m=1}^M\Pr\left\{|\boldX_j^{m \rm \top}W_m\boldepsilon^m| > \frac{np'_\lambda(0+)}{(1+\psi_m)}\right\}\nonumber\\
&\leq& 2p \sum\limits_{m=1}^M\exp\left\{-\frac{n^2p'^2_\lambda(0+) }{2n_m \Lambda_m \sigma_m^2(1+\psi_m)^2}\right\}.
\eqe
Part (2) is proved by combining (\ref{kkt1:homo}), (\ref{kkt2:homo}), (\ref{tt1}), and (\ref{tt2}). \hfill $\Box$

\vspace{1cm}

\noindent{\bf Proof of Theorem 3.}
The proof is similar to that of Part 1 of Theorem 2 and is omitted here.   \hfill $\Box$

\vspace{1cm}

\noindent{\bf Proof of Theorem 4.}
By the Karush-Kuhn-Tucher(KKT) conditions, we need to prove that $\check{\bbeta}$ satisfies
\bqe
-X_{S_m}^{m \rm \top}\left(\boldy^m-X_{S_m}^{m}\check{\bbeta}_{S_m}^m\right)+n p'_{O,\lambda_O}(\sum_{m=1}^M p_{I,\lambda_I}(|\check{\bbeta}_{S_m}^m|))\circ p'_{I,\lambda_I}(|\check{\bbeta}_{S_m}^m|)=0,\label{kkt1:heter}\\
|X_{S-S_m}^{m \rm \top}(\boldy^m- X_{S_m}^m\check{\bbeta}_{S_m}^m)|\leq np'_{I,\lambda_I}(0+)p'_{O,\lambda_O}(\sum_{m=1}^M p_{I,\lambda_I}(|\check{\bbeta}_{S-S_m}^m|)),~~~~~ \label{kkt2:heter}\\
\|X_{S^c}^{m \rm \top}(\boldy^m- X_{S_m}^m\check{\bbeta}_{S_m}^m)\|_\infty\leq np'_{O,\lambda_O}(0+)p'_{I,\lambda_I}(0+). ~~~~~~~~~~~~~~~~~~~~~~~~\label{kkt3:heter}
\eqe

If $\min\limits_{j \in S_m} |\check{\beta}_j^m|>\theta_I\lambda_I$, then $p'_{I,\lambda_I}(|\check{\bbeta}_{S_m}^m|)=0$. Recall the definition of the estimator $\check{\bbeta}_{S_m}^m$. We can easily get (\ref{kkt1:heter}). Set $$C_m^\dag \leq \frac{\min\limits_{(j,m)\in \mathcal {A}}|{\beta_j^m}^*|}{2}\sqrt{\frac{n_m}{|S_m|}}.$$
Note that $\lambda_I < \frac{\min\limits_{(j,m)\in \mathcal {A}}|{\beta_j^m}^*|}{2\theta_I}$. Therefore,
\bqe\label{bound1}
&&\Pr\left\{\min\limits_{(j,m)\in \mathcal {A}} |\check{\beta}_j^m|>\theta_I\lambda_I\right\}\geq \Pr\left(\|\check{\bbeta}_{S_m}^{ m} - {\bbeta_{S_m}^{m}}^*\|_2\leq \sqrt{\frac{|S_m|}{n_m}}C_m^\dag ,~m=1,\cdots,M\right)\nonumber\\
&\geq& 1- 2\sum\limits_{m = 1}^M \exp\left\{-C_m^{\dag 2}\frac{|S_m|(\underline{\rho}_1^{*m})^2}{8\bar{\rho}_{2}^{*m}\sigma_m^2 }\right\}.
\eqe
In fact,
\[
X_{S-S^m}^{m \rm \top}(\boldy^m- X_{S_m}^m\check{\bbeta}_{S_m}^m)=\boldX_{S-S^m}^{m \rm \top}W_m\boldepsilon^m -\boldX_{S-S^m}^{m \rm \top}W_m X_{S_m}^m(\check{\bbeta}_{S_m}^m -{\bbeta_{S_m}^m}^*),
\]
and $\check{\bbeta}_{S_m}^m -{\bbeta_{S_m}^m}^*=\left(\boldX_{S_m}^{m \rm \top}W_m \boldX_{S_m}^{m}\right)^{-1}X_{S_m}^{m \rm \top}W_m \boldepsilon^m$.
Then we have
\bqe\label{skkt2}
&&|X_{S-S_m}^{m \rm \top}(\boldy^m- X_{S_m}^m\check{\bbeta}_{S_m}^m)| \nonumber\\
&\leq& |\boldX_{S-S^m}^{m \rm \top}W_m\boldepsilon^m | + |\boldX_{S-S^m}^{m \rm \top}W_m X_{S_m}^m \left(\boldX_{S_m}^{m \rm \top}W_m \boldX_{S_m}^{m}\right)^{-1}X_{S_m}^{m \rm \top}W_m \boldepsilon^m| \nonumber\\
&\leq& |\boldX_{S-S^m}^{m \rm \top}W_m\boldepsilon^m  | + \left\|\boldX_{S-S^m}^{m \rm \top}W_m X_{S_m}^m \left(\boldX_{S_m}^{m \rm \top}W_m \boldX_{S_m}^{m}\right)^{-1}\right\|_\infty |X_{S_m}^{m \rm \top}W_m \boldepsilon^m|\nonumber\\
&\leq& \left\|\boldX_{S}^{m \rm \top}W_m \boldepsilon^m\right\|_\infty + \psi_m^*\left\|\boldX_{S}^{m \rm \top}W_m \boldepsilon^m\right\|_\infty \nonumber\\
 &\leq&
\left\|\boldX_S^{m \rm \top}W_n \boldepsilon^m\right\|_\infty (1+\psi_m^*).
\eqe
Hence (\ref{kkt2:heter}) holds when
\bqe\label{kkt2c}
\left\|\boldX_S^{m \rm \top}W_m \boldepsilon^m\right\|_\infty \leq  \frac{np'_{I,\lambda_I}(0+)p'_{O,\lambda_O}(J^{-m} f_I^{max})}{(1+\psi_m^*)}.
\eqe
That is because for $m=1,\cdots,M$,
\bse
&&|X_{S-S_m}^{m \rm \top}(\boldy^m- X_{S_m}^m\check{\bbeta}_{S_m}^m)|\leq \frac{np'_{I,\lambda_I}(0+)p'_{O,\lambda_O}(J^{-m} f_I^{max})}{(1+\psi_m^*)}(1+\psi_m^*) \\
&\leq& np'_{I,\lambda_I}(0+)p'_{O,\lambda_O}(J^{-m} f^{max}_I)}{(1+\psi_m^*)\\
&\leq& np'_{I,\lambda_I}(0+)p'_{O,\lambda_O}\left(\sum_{m=1}^M p_{I,\lambda_I}(|\check{\bbeta}_{S-S_m}^m|)\right).
\ese
We now derive the probability bounds for the event in (\ref{kkt2c}). In fact, by Bonferroni's inequality and sub-Gaussian tail probability bound in Condition 1,
\bqe\label{bound2}
&&\Pr\left\{\left\|\boldX_S^{m \rm \top}W_m \boldepsilon^m\right\|_\infty> \frac{np'_{I,\lambda_I}(0+)p'_{O,\lambda_O}(J^{-m} f_I^{max})}{(1+\psi_m^*)},~ \exists~m \in \{1,\cdots,M\}\right\}\nonumber\\
&\leq& 2|S| \sum\limits_{m=1}^M\exp\left\{-\frac{n^2 p'^2_{I,\lambda_I}(0+)p'^2_{O,\lambda_O}(J^{-m} f_I^{max})}{2n_m \bar{\rho}_2^{*m}\sigma_m^2(1+\psi^*_m)^2}\right\}.
\eqe
Similarly, we can prove (\ref{kkt3:heter}). Actually,
\bse
&&|X_{S^c}^{m \rm \top}(\boldy^m- X_{S_m}^m\check{\bbeta}_{S_m}^m)|_\infty \leq \left\|\boldX_{S^c}^{m \rm \top}W_m \boldepsilon^m\right\|_\infty + \psi_m^*\left\|\boldX_{S}^{m \rm \top}W_m \boldepsilon^m\right\|_\infty \\
&<&
\left\|\boldX_{S^c}^{m \rm \top}W_m\boldepsilon^m\right\|_\infty + \frac{\psi_m^*}{(1+\psi_m^*)}np'_{O,\lambda_O}(0+)p'_{I,\lambda_I}(0+).
\ese
Based on the above discussions, (\ref{kkt3:heter}) follows if $\left\|\boldX_{S^c}^{m \rm \top}W_m \boldepsilon^m\right\|_\infty<\frac{np'_{O,\lambda_O}(0+)p'_{I,\lambda_I}(0+)}
{(1+\psi_m^*)}$. The probability bound is derived as
\bqe\label{bound3}
&&\Pr\left\{\left\|\boldX_{S^c}^{m \rm \top}W_m \boldepsilon^m\right\|_\infty>\frac{np'_{O,\lambda_O}(0+)p'_{I,\lambda_I}(0+)}
{(1+\psi_m^*)},~ \exists~m \in \{1,\cdots,M\}\right\}\nonumber\\
&\leq & 2 (p-|S|) \sum\limits_{m=1}^M\exp\left\{-\frac{n^2 p'^2_{I,\lambda_I}(0+)p'^2_{O,\lambda_O}(0+)}{2n_m \Lambda_m\sigma_m^2(1+\psi^*_m)^2}\right\}.
\eqe
Therefore, the theorem is proved by combining (\ref{kkt1:heter}), (\ref{kkt2:heter}), (\ref{kkt2:heter}), (\ref{bound1}),(\ref{bound2})  and (\ref{bound3}) . \hfill $\Box$

\vspace{1cm}

\noindent{\bf Proof of Theorem 5.} By the Karush-Kuhn-Tucher(KKT) conditions, we need to prove that $\check{\bbeta}$ satisfies
\bqe
&&-X_{S_m}^{m \rm \top}\left(\boldy^m-X_{S_m}^{m}\check{\bbeta}_{S_m}^m\right)
        +n p'_{1,\lambda_1}(\|\check{\bbeta}_{S_m}\|_2)
        \circ\frac{\check{\bbeta}_{S_m}^m}{\|\check{\bbeta}_{S_m}\|_2}\nonumber\\
&&~~~~~~~~+n p'_{2,\lambda_2}(|\check{\bbeta}_{S_m}^m|)\circ \mbox{sgn}(\check{\bbeta}_{S_m}^m)=0,\label{kkt1:heter2}\\
&&\|X_{S-S_m}^{m \rm \top}(\boldy^m- X_{S_m}^m\check{\bbeta}_{S_m}^m)\|_\infty\leq np'_{2,\lambda_2}(0+), \label{kkt2:heter2}\\
&&\|X_{S^c}^{m \rm \top}(\boldy^m- X_{S_m}^m\check{\bbeta}_{S_m}^m)\|_\infty\leq np'_{1,\lambda_1}(0+)+np'_{2,\lambda_2}(0+). \label{kkt3:heter2}
\eqe

Note that $\check{\bbeta}_{S_m}^m$ satisfies $-X_{S_m}^{m \rm \top}\left(\boldy^m-X_{S_m}^{m}\check{\bbeta}_{S_m}^m\right)=0$.
If $$\min\limits_{j \in S_m} |\check{\beta}_j^m|>\theta_2\lambda_2 ~\mbox{ and }\min\limits_{j \in S} \|\check{\bbeta}_j\|_2>\theta_1\lambda_1,$$ then we have (\ref{kkt1:heter2}). Set $C_m^\dag \leq \frac{\min\limits_{(j,m)\in \mathcal {A}}|{\beta_j^m}^*|}{2}\sqrt{\frac{n_m}{|S_m|}}.$
Note that $$\lambda_1 < \frac{\min\limits_{j \in S}\|{\bbeta_j}^*\|_2}{2\theta_1},~\lambda_2 < \frac{\min\limits_{(j,m)\in \mathcal {A}}|{\beta_j^m}^*|}{2\theta_2}.$$ Therefore, if $\min\limits_{(j,m)\in \mathcal {A}} |\check{\beta}_j^m|>\theta_2\lambda_2$, then we must have
$\min\limits_{j \in S} \|\check{\bbeta}_j\|_2>\theta_1\lambda_1$.
\bqe\label{abd1}
&&\Pr\left\{\min\limits_{(j,m)\in \mathcal {A}} |\check{\beta}_j^m|>\theta_1\lambda_1, ~\min\limits_{j \in S} \|\check{\bbeta}_j\|_2>\theta_2\lambda_2\right\}\nonumber\\
&\geq & \Pr\left(\|\check{\bbeta}_{S_m}^{ m} - {\bbeta_{S_m}^{m}}^*\|_2\leq \sqrt{\frac{|S_m|}{n_m}}C_m^\dag ,~m=1,\cdots,M\right)\nonumber\\
&\geq& 1- 2\sum\limits_{m = 1}^M \exp\left\{-C_m^{\dag 2}\frac{|S_m|(\underline{\rho}_{1}^{* m})^2}{8\bar{\rho}_{2}^{* m}\sigma_m^2 }\right\}.
\eqe
Similar as the proof of Theorem 4, (\ref{kkt2:heter2}) holds when
\bqe\label{kkt2ca}
\left\|\boldX_S^{m \rm \top}W_m \boldepsilon^m\right\|_\infty \leq  \frac{np'_{2,\lambda_2}(0+)}{(1+\psi_m^*)}.
\eqe
Then we have
\bqe\label{abd2}
&&\Pr\left\{\left\|\boldX_S^{m \rm \top}W_m \boldepsilon^m\right\|_\infty> \frac{np'_{2,\lambda_2}(0+)}{(1+\psi_m^*)},~ \exists~m \in \{1,\cdots,M\}\right\}\nonumber\\
&\leq& 2|S| \sum\limits_{m=1}^M\exp\left\{-\frac{n^2 p'^2_{2,\lambda_2}(0+)}{2n_m \bar{\rho}_{2}^{* m}\sigma_m^2(1+\psi^*_m)^2}\right\}.
\eqe
Similarly, we can show (\ref{kkt3:heter2}) holds when $\left\|\boldX_{S^c}^{m \rm \top}W_m \boldepsilon^m\right\|_\infty<\frac{np'_{1,\lambda_1}(0+)+np'_{2,\lambda_2}(0+)}
{(1+\psi_m^*)}$. The probability bound is derived as
\bqe\label{abd3}
&&\Pr\left\{\left\|\boldX_{S^c}^{m \rm \top}W_m \boldepsilon^m\right\|_\infty>\frac{np'_{1,\lambda_1}(0+)+np'_{2,\lambda_2}(0+)}
{(1+\psi_m^*)},~ \exists~m \in \{1,\cdots,M\}\right\}\nonumber\\
&\leq & 2 (p-|S|) \sum\limits_{m=1}^M\exp\left\{-\frac{n^2 [p'_{1,\lambda_1}(0+)+p'_{2,\lambda_2}(0+)]^2}{2n_m \Lambda_m\sigma_m^2(1+\psi^*_m)^2}\right\}.
\eqe
Therefore, the theorem is proved by combining (\ref{kkt1:heter2}), (\ref{kkt2:heter2}), (\ref{kkt2:heter2}), (\ref{abd1}), (\ref{abd2}),  and (\ref{abd3}). \hfill $\Box$

\clearpage
\section{Additional Numerical Results}

\begin{table}[htbp]
\caption{Analysis of lung cancer data using SGMCP: identified genes and their estimates.}
\centering
{
\begin{tabular}{cccccc}
\hline
Probe & Gene &  UM & HLM & DFCI & MSKCC  \\
\hline
201462$\_$at	&	SCRN1	&		&	0.0034	&		&	0.0020	\\
202831$\_$at	&	GPX2	&		&	-0.0022	&		&	-0.0021	\\
203917$\_$at	&	CXADR	&	0.0021	&		&	0.0004	&	0.0066	\\
205776$\_$at	&	FMO5	&	0.0005	&	0.0035	&	0.0038	&		\\
206754$\_$s$\_$at	&	CYP2B6	&	0.0012	&		&	0.0020	&		\\
207850$\_$at	&	CXCL3	&		&	-0.0216	&		&	0.0120	\\
208025$\_$s$\_$at	&	HMGA2	&	-0.0028	&	0.0001	&	-0.0037	&	-0.0012	\\
219654$\_$at	&	PTPLA	&	-0.0025	&	-0.0145	&		&	0.0055	\\
219764$\_$at	&	FZD10	&	-0.0005	&	-0.0019	&	-1.6E-05	&	-0.0022	\\
\hline
\end{tabular}
}
\end{table}

\begin{table}
\caption{ Analysis of lung cancer data using meta-analysis: identified genes and their estimates.}
\centering
{
\begin{tabular}{cccccc}
\hline
Probe & Gene &  UM & HLM & DFCI & MSKCC  \\
\hline
201462$\_$at	&	SCRN1	&		&	0.0101	&		&		\\
203559$\_$s$\_$at	&	ABP1	&		&	0.0005	&		&		\\
203876$\_$s$\_$at	&	MMP11	&		&		&		&	-0.0066	\\
203921$\_$at	&	CHST2	&	0.0051	&		&		&		\\
204855$\_$at	&	SERPINB5	&	-0.0012	&		&		&		\\
206754$\_$s$\_$at	&	CYP2B6	&		&		&	0.0104	&		\\
206994$\_$at	&	CST4	&		&	-0.0037	&		&		\\
207850$\_$at	&	CXCL3	&		&	-0.0246	&		&		\\
208025$\_$s$\_$at	&	HMGA2	&	-0.0021	&		&	-0.0010	&		\\
209343$\_$at	&	EFHD1	&		&		&		&	0.0096	\\
212328$\_$at	&	LIMCH1	&		&		&	0.0050	&		\\
213703$\_$at	&	LINC00342	&	0.0008	&		&		&		\\
215867$\_$x$\_$at	&	CA12	&		&	-0.0026	&		&		\\
218677$\_$at	&	S100A14	&		&		&		&	-0.0257	\\
218824$\_$at	&	PNMAL1	&	0.0003	&		&		&		\\
219654$\_$at	&	PTPLA	&		&	-0.0240	&		&		\\
219747$\_$at	&	NDNF	&	0.0002	&		&		&		\\
220952$\_$s$\_$at	&	PLEKHA5	&		&	-0.0047	&		&		\\
221841$\_$s$\_$at	&	KLF4	&	-0.0047	&		&		&		\\
222043$\_$at	&	CLU	&		&	0.0049	&		&		\\

\hline
\end{tabular}
}
\end{table}

\begin{table}
\caption{Analysis of lung cancer data using pooled analysis: identified genes and their estimates.}
\centering
{
\begin{tabular}{cccccc}
\hline
Probe & Gene &  UM & HLM & DFCI & MSKCC  \\
\hline
201462$\_$at	&	SCRN1	&		&	0.0101	&		&		\\
203559$\_$s$\_$at	&	ABP1	&		&	0.0005	&		&		\\
203876$\_$s$\_$at	&	MMP11	&		&		&		&	-0.0066	\\
203921$\_$at	&	CHST2	&	0.0051	&		&		&		\\
204855$\_$at	&	SERPINB5	&	-0.0012	&		&		&		\\
206754$\_$s$\_$at	&	CYP2B6	&		&		&	0.0104	&		\\
206994$\_$at	&	CST4	&		&	-0.0037	&		&		\\
207850$\_$at	&	CXCL3	&		&	-0.0246	&		&		\\
208025$\_$s$\_$at	&	HMGA2	&	-0.0021	&		&	-0.0010	&		\\
209343$\_$at	&	EFHD1	&		&		&		&	0.0096	\\
212328$\_$at	&	LIMCH1	&		&		&	0.0050	&		\\
213703$\_$at	&	LINC00342	&	0.0008	&		&		&		\\
215867$\_$x$\_$at	&	CA12	&		&	-0.0026	&		&		\\
218677$\_$at	&	S100A14	&		&		&		&	-0.0257	\\
218824$\_$at	&	PNMAL1	&	0.0003	&		&		&		\\
219654$\_$at	&	PTPLA	&		&	-0.0240	&		&		\\
219747$\_$at	&	NDNF	&	0.0002	&		&		&		\\
220952$\_$s$\_$at	&	PLEKHA5	&		&	-0.0047	&		&		\\
221841$\_$s$\_$at	&	KLF4	&	-0.0047	&		&		&		\\
222043$\_$at	&	CLU	&		&	0.0049	&		&		\\
\hline
\end{tabular}
}
\end{table}

\begin{table}
\caption{Analysis of lung cancer data using GMCP: identified genes and their estimates.}
\centering
{
\begin{tabular}{cccccc}
\hline
Probe & Gene &  UM & HLM & DFCI & MSKCC  \\
\hline
202503$\_$s$\_$at	&	KIAA0101	&	-0.0009	&	-0.0020	&	-0.0021	&	-0.0019	 \\
205776$\_$at	&	FMO5	&	0.0001	&	0.0002	&	0.0002	&	-0.0001	\\
207850$\_$at	&	CXCL3	&	-0.0017	&	-0.0139	&	0.0029	&	0.0095	\\
208025$\_$s$\_$at	&	HMGA2	&	-3.2E-05	&	1.1E-05	&	-3.8E-05	&	 -2.2E-05	\\
219654$\_$at	&	PTPLA	&	-0.0036	&	-0.0092	&	-0.0024	&	0.0060	\\
219764$\_$at	&	FZD10	&	-0.0014	&	-0.0036	&	-0.0014	&	-0.0036	\\
\hline
\end{tabular}
}
\end{table}

\end{spacing}

\end{document}